\documentclass{iaus}
\usepackage{graphicx}
\AtBeginDvi{}

\title[LC Models for SN 2009dc] 
{Light Curve Models for SN 2009dc}

\author[Y. Kamiya]   
{Yasuomi Kamiya
}

\affiliation{
Department of Astronomy, Graduate School of Science, the University of Tokyo,\\
7-3-1 Hongo, Bunkyo-ku, Tokyo 113-0033, Japan\\[\affilskip]
Institute for the Physics and Mathematics of the Universe,\\
Todai Institutes for Advanced Study, the University of Tokyo,\\
5-1-5 Kashiwanoha, Kashiwa, Chiba 277-8583, Japan\\[\affilskip]
email: {\tt yasuomi.kamiya@ipmu.jp}
}

\pubyear{2011}
\volume{281}  
\pagerange{}
\jname{Binary Paths to Type Ia Supernova Explosions}
\editors{Rosanne Di Stefano \& Marina Orio, eds.}

\newcommand{\Ms}{M_\odot}
\newcommand{\Mni}{M_\mathrm{Ni}}
\newcommand{\Mwd}{M_\mathrm{WD}}
\newcommand{\Mece}{M_\mathrm{ECE}}
\newcommand{\Mime}{M_\mathrm{IME}}
\newcommand{\Mco}{M_\mathrm{CO}}
\newcommand{\Ekin}{E_\mathrm{kin}}
\newcommand\ion[2]{#1~\textsc{#2}}
\newcommand{\uvoir}{\textit{uvoir}}
\newcommand{\BVRI}{\textit{BVRI}}
\newcommand{\vph}{v_\mathrm{ph}}

\newcommand{\apjl}{\textit{ApJ} (Letters)}
\newcommand{\apj}{\textit{ApJ}}
\newcommand{\mnras}{\textit{MNRAS}}
\newcommand{\aapl}{\textit{A\&A} (Letters)}
\newcommand{\aap}{\textit{A\&A}}

\begin{document}

\maketitle

\begin{abstract}
Simplified explosion models of super-Chandrasekhar-mass C-O white dwarfs (WDs) are constructed with parameters such as WD mass and $^{56}$Ni mass.
Their light curves are obtained by solving one-dimensional equations of radiation hydrodynamics, and compared with the observations of SN 2009dc, one of the overluminous Type Ia supernovae, to estimate its properties.
As a result, the progenitor of SN 2009dc is suggested to be a 2.2--2.4-$\Ms$ C-O WD with 1.2--1.4 $\Ms$ of $^{56}$Ni, if the extinction by its host galaxy is negligible.
\end{abstract}


SN 2009dc is one of the overluminous Type Ia supernovae (SNe Ia) and estimated to have $\geq$1.2 $\Ms$ of $^{56}$Ni (e.g., \cite[Yamanaka \etal\ 2009]{Yamanaka09}).
To explain the production of such a large mass of $^{56}$Ni, its progenitor is  proposed to be a super-Chandrasekhar-mass (super-Ch-mass) C-O white dwarf (WD).
An asymmetric explosion of a Ch-mass C-O WD could also explain the high luminosity of SN 2009dc (\cite[Hillebrandt \etal\ 2007]{HIllebrandt07}), which is unlikely because of the observed small polarization for SN 2009dc (\cite[Tanaka \etal\ 2010]{Tanaka10}).

To study the properties of SN 2009dc from its light curves (LCs), simplified explosion models of super-Ch-mass C-O WDs are constructed in a manner almost similar to \cite{Maeda09}.
A model is parameterized by its masses of WD, electron-captured (stable Fe-peak) elements (ECEs; Fe, Co, and $^{58}$Ni), $^{56}$Ni, intermediate-mass elements (IMEs; Si, S, Ca), and C and O ($\Mwd$, $\Mece$, $\Mni$, $\Mime$, and $\Mco$, respectably).
Once the above parameters are set, its kinetic energy ($\Ekin$) is calculated, assuming that the central density is $3\times10^9$ g cm$^{-3}$.
Then the density and velocity structures are obtained by scaling those of W7 (\cite[Nomoto \etal\ 1984, Thielemann \etal\ 1986]{Nomoto84,Thielemann86}) with $\Mwd$ and $\Ekin$.
For its abundance distribution, the model consists of ECEs, $^{56}$Ni, IMEs, and C-O layers, from the center to the surface.
However, mixing is assumed for a certain region due to the observed slow \ion{Si}{ii} line velocity (\cite[Yamanaka \etal\ 2009]{Yamanaka09}).
The parameter ranges in this study are $\Mwd/\Ms=1.8$, 2.0, $\dots$, 2.8; $\Mni/\Ms=1.2$, 1.4, 1.6, 1.8; $\Mece/\Mwd=0.1$, 0.2, $\dots$; and $\Mco/\Mwd=0.1$, 0.2, $\dots$.
With the relation, $\Mwd=\Mece+\Mni+\Mime+\Mco$, $\Mime$ is hereafter not indicated.\@
Any model whose $\Ekin$ is not positive is excluded.

STELLA code is used to calculate their LCs, which solves one-dimensional equations of radiation hydrodynamics (e.g., \cite[Blinnikov \etal\ 1998]{Blinnikov98}).
In this study, multi-band ($U$-, $B$-, $V$-, $R$-, and $I$-band) LCs are obtained, as well as
bolometric ones.
The ``bolometric'' LCs in the observations are not truly bolometric, but \uvoir.
For SN 2009dc, the observed \uvoir\ luminosity is assumed to be $1/(60\%)$ of the integrated \BVRI\ one (\cite[Yamanaka \etal\ 2009]{Yamanaka09}), which is common for normal SNe Ia (\cite[Wang \etal\ 2009]{Wang09}).
Since it is still unknown whether this assumption is true for overluminous SNe Ia, \BVRI\ LCs are compared in the following.

\begin{figure}
\begin{center}
\includegraphics[width=0.45\linewidth]{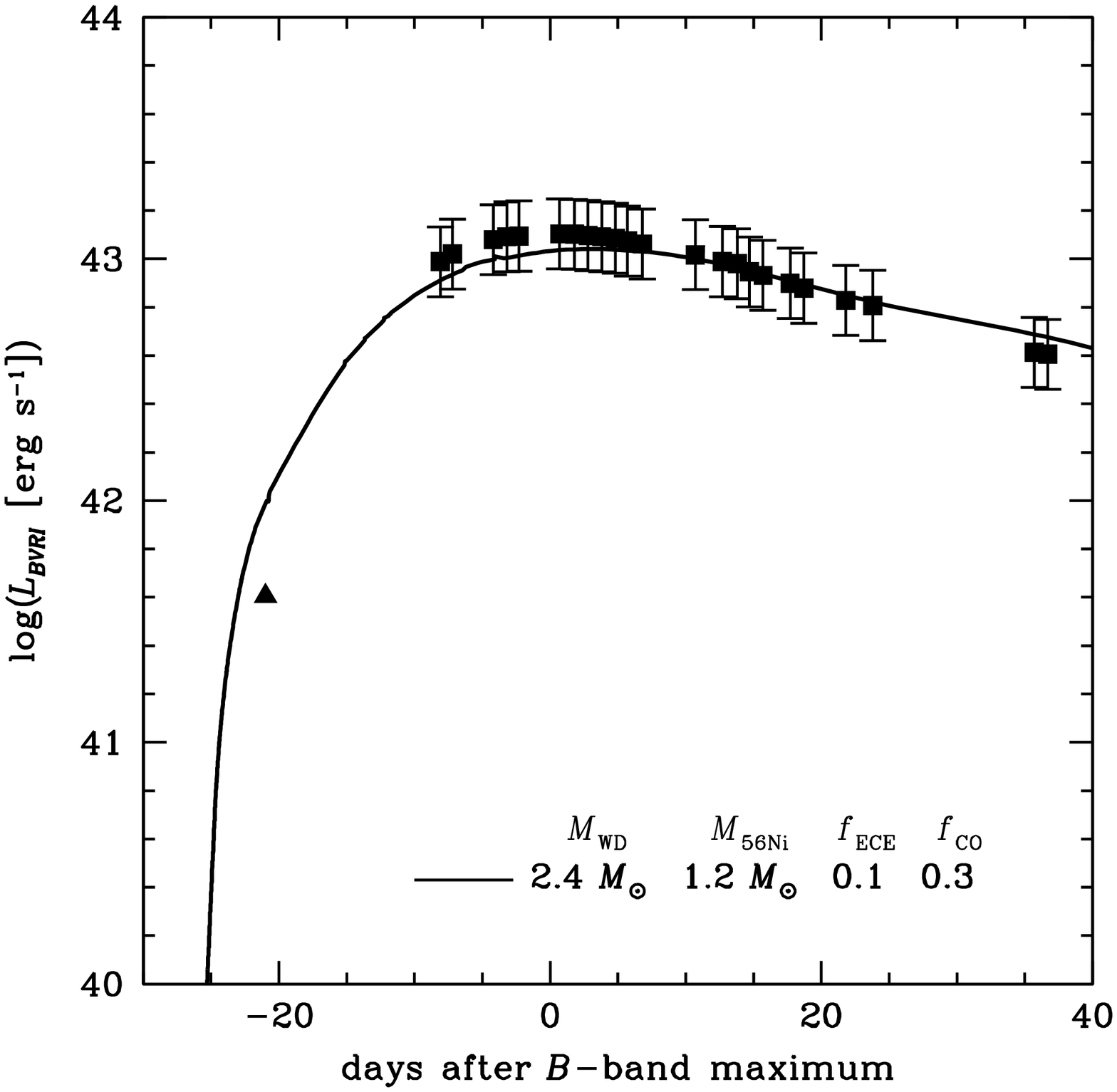}
\includegraphics[width=0.45\linewidth]{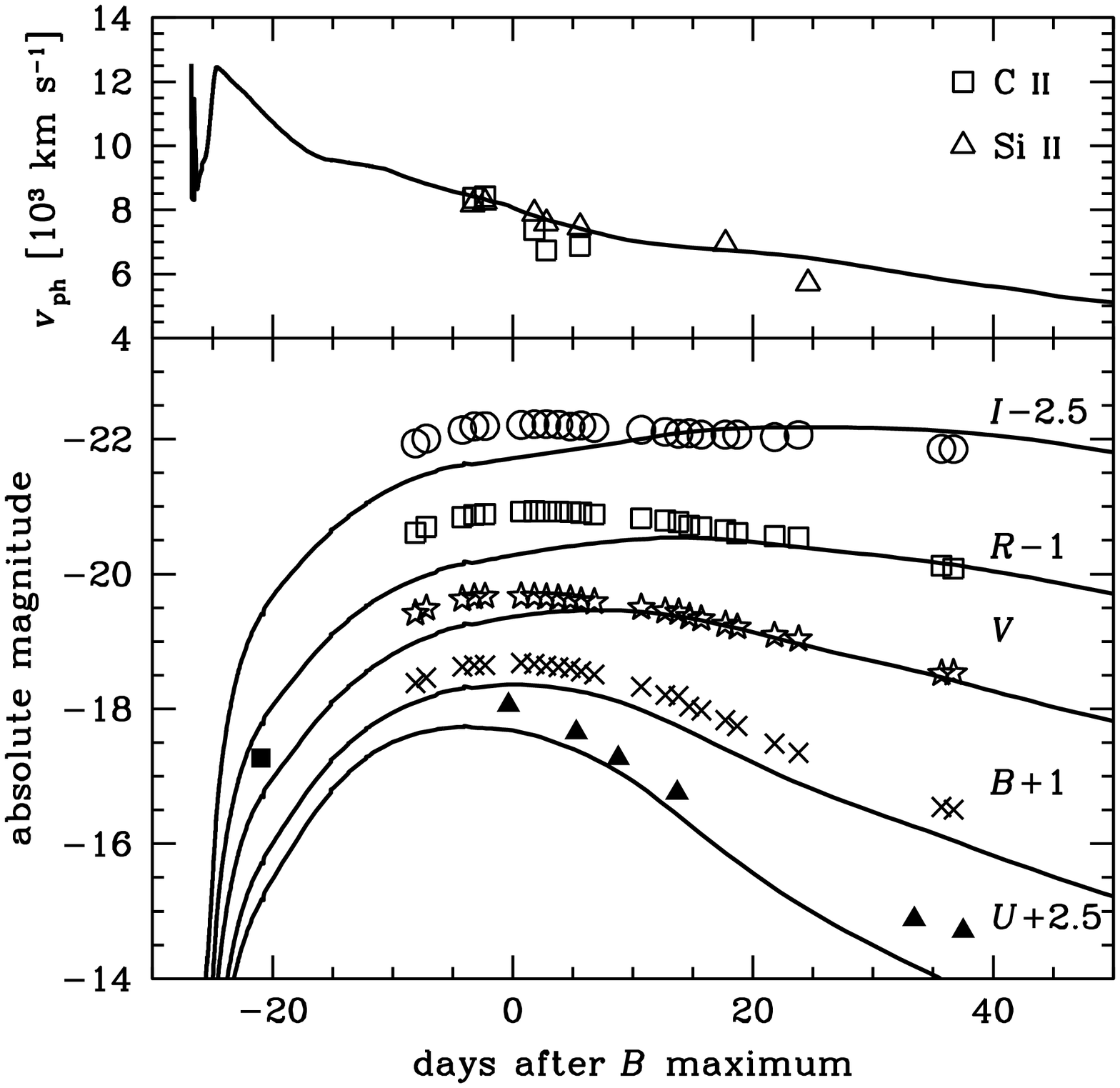}
\caption{Comparison of the most preferable model with SN 2009dc.
The model parameters are indicated in the left panel.
The observation data are taken from \cite{Yamanaka09}, except for the early detection (triangle in the left panel, filled square in the right-bottom), estimated from \cite{Silverman11}.
Note that the extinction by the host galaxy is neglected.}
\label{fig1}
\end{center}
\end{figure}

The resultant \BVRI\ LCs of the models show the similar tendency as in \cite{Maeda09}; a massive WD model has a wider \BVRI\ LC and vice versa.
Since SN 2009dc shows a wide \BVRI\ LC, lighter WD models such as $\Mwd=$ 1.8 $\Ms$ can be excluded.
Many models have much larger photospheric velocity ($\vph$) than the observed line velocity.
To make a further comparison between the \BVRI\ LCs of the models and the observations without the host-galaxy extinction, $\chi^2$ is calculated for the models with smaller $\vph$.
Among them, the models with $\Mwd=2.2$--2.4 $\Ms$, $\Mni=1.2$--1.4 $\Ms$, $\Mece=0.1\Mwd$, and $\Mco=0.3\Mwd$ show relatively smaller $\chi^2$, being well fitted to the observation (Figure \ref{fig1}).
These models indicate that the progenitor of SN 2009dc should have a relatively massive C-O layer, which results in smaller $\Ekin$ and thus $\vph$.

\begin{acknowledgments}

Y.K. is grateful to his collaborators, Masaomi Tanaka, Ken'ichi Nomoto, Sergei I. Blinnikov, Elena I. Sorokina, and Tomoharu Suzuki.
As a research fellow of the Japan Society for the Promotion of Science (JSPS), he and his work are financially supported by the JSPS Research Fellowships for Young Scientists and Grant-in-Aid for JSPS Fellows \#22$\cdot$7641, respectably.

\end{acknowledgments}

%


\begin{thebibliography}{}

\bibitem[Blinnikov \etal\ (1998)]{Blinnikov98}
{Blinnikov, S. I., Eastman, R., Bartunov, O. S., \etal} 1998, \apj, 496, 454

\bibitem[Hillebrandt \etal\ (2007)]{Hillebrandt07}
{Hillebrandt, W., Sim, S. A., \& R\"{o}pke, F. K.} 2007, \aapl, 465, L17

\bibitem[Maeda \& Iwamoto (2009)]{Maeda09}
{Maeda, K., \& Iwamoto, K.} 2009, \mnras, 394, 239

\bibitem[Nomoto \etal\ (1984)]{Nomoto84}
{Nomoto, K., Thielemann, F.-K., \& Yokoi, K.} 1984, \apj, 286, 644

\bibitem[Silverman \etal\ (2011)]{Silverman11}
{Silverman, J. M., Ganeshalingam, M., Li, W., \etal} 2011, \mnras, 410, 585

\bibitem[Tanaka \etal\ (2010)]{Tanaka10}
{Tanaka M., Kawabata K. S., Yamanaka M., \etal} 2010, \apj, 714, 1209

\bibitem[Thielemann \etal\ (1986)]{Thielemann86}
{Thielemann, F.-K., Nomoto, K., \& Yokoi, K.} 1986, \aap, 158, 17

\bibitem[Wang \etal\ (2009)]{Wang09}
{Wang, X., Li, W., Filippenko, A. V., \etal} 2009, \apj, 697, 380

\bibitem[Yamanaka \etal\ (2009)]{Yamanaka09}
{Yamanaka M., Kawabata K. S., Kinugasa K., \etal} 2009, \apjl, 707, L118

\end{thebibliography}
\end{document}